\documentclass[twocolumn,preprintnumbers,amsmath,amssymb]{revtex4}
\usepackage{graphicx}
\usepackage{dcolumn}
\usepackage{bm}
\usepackage{hyperref} 
\usepackage{multirow} 
\usepackage{soul}
\newcommand\beq{\begin{equation}}
\newcommand\eeq{\end{equation}}
\newcommand\bea{\begin{eqnarray}}
\newcommand\eea{\end{eqnarray}}
\usepackage{wrapfig}

\begin{document}

\title{\bf Density matrix renormalization group algorithm for Bethe lattices of spin 1/2 or 1 sites with Heisenberg antiferromagnetic exchange \\}
\author{\bf Manoranjan Kumar$^{1,2}$, S. Ramasesha$^2$ and Zolt\'an G. Soos$^1$ }
\address{ $^{1}$Department of Chemistry, Princeton University, Princeton New Jersey 08544 \\}
\address{ $^2$Solid State and Structural Chemistry Unit, Indian Institute of Science, Bangalore 560012, India.\\}
\date{\today}

\begin{abstract}

An efficient density matrix renormalization group (DMRG) algorithm is 
presented for the Bethe lattice with connectivity $Z = 3$ and antiferromagnetic exchange 
between nearest neighbor spins $s= 1/2$ or 1 sites in successive generations $g$. 
The algorithm is accurate for $s = 1$ sites. The ground states are magnetic 
with spin $S(g) = 2^g s$, staggered magnetization that persists for 
large $g > 20$ and short-range spin correlation functions that decrease exponentially. A finite energy 
gap to $S > S(g)$ leads to a magnetization plateau in the extended lattice. 
Closely similar DMRG results for $s$ = 1/2 and 1 are interpreted in terms of an analytical three-site model.  
\vskip .4 true cm
\noindent PACS numbers: 71.10.Fd, 75.10.Jm, 05.30.-d, 05.10.Cc  \\
\noindent Email:manoranj@princeton.edu
\end{abstract}
\maketitle

\section{Introduction}
\begin{figure}
\begin{center}
\includegraphics[height=8.0cm,width=9.0cm]{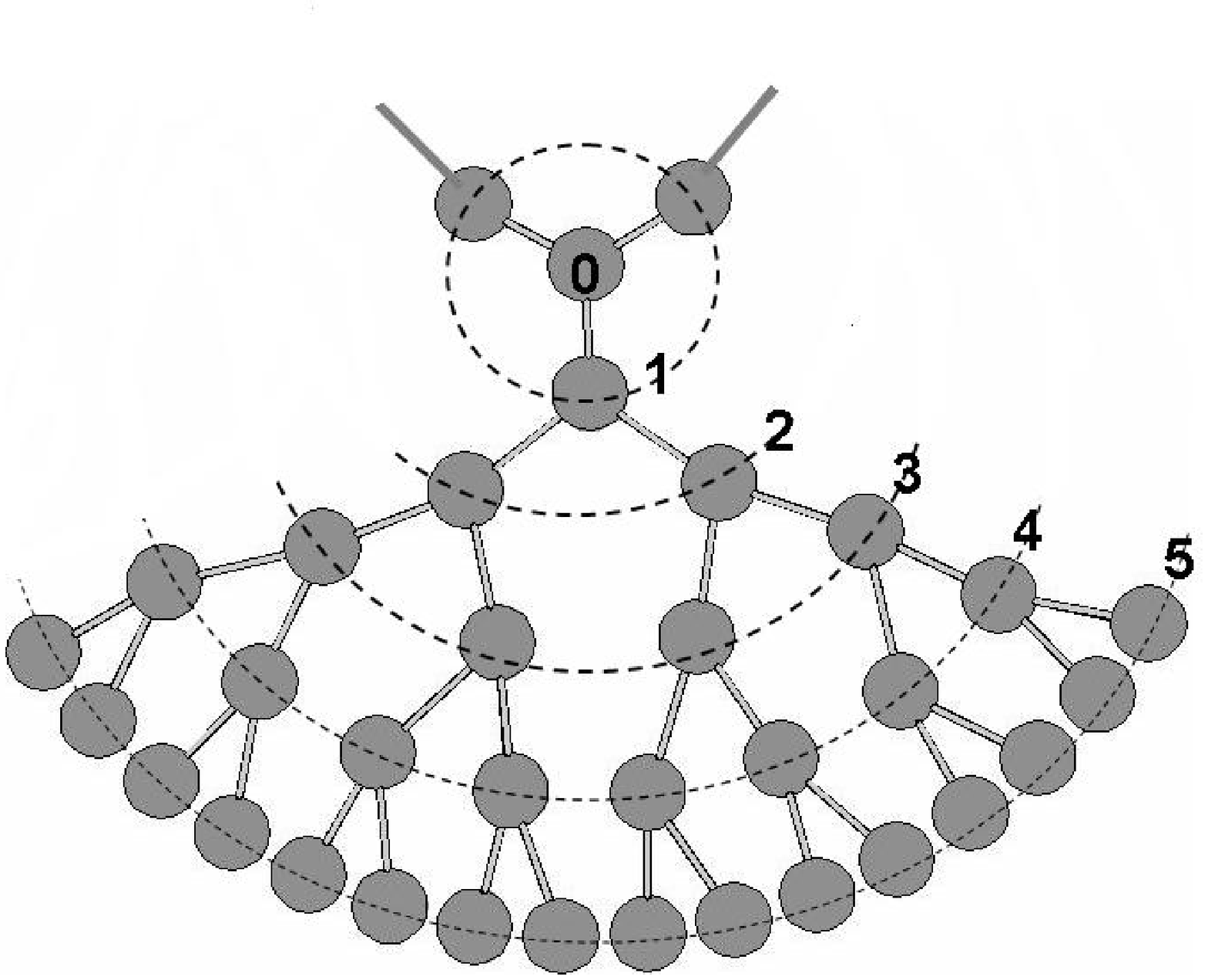}
\caption{ One arm of a Bethe lattice with $b = Z = 3$ and $g = 5$. 
The other two arms are shown to generation $g = 1$; $g$ also 
represents distance from the focal point.}
\label{fig1}
\end{center}
\end{figure}

Dendrimers have been extensively explored over the past two decades, 
both experimentally and theoretically. Dendrimer are macromolecules with 
repetitive branches attached to a central core or focal point \cite{r1,r2}. 
Three parameters (b,Z,g) characterize the dendrimer lattice model in 
Fig.\ref{fig1}: the number of branches b attached to the focal point, 
the connectivity $Z$ of a site and the number of generations $g$. Many natural 
systems are dendrimers and molecular engineering can produce custom dendrimers with 
constituents that range from carbon-based molecules \cite{r3} to organo-metallic 
compounds \cite{r4,r5}. Potential applications include drug delivery for 
chemotherapy \cite{r3}, gene therapy \cite{r6}, magnetic resonance imaging as 
contrast agents based on superparamagnetism \cite{r7}, and molecular recognition \cite{r8}.\\

Bethe lattices (BL) or Cayley trees are theoretical models of dendrimers and attractive 
approximations of solid-state systems. We consider in this paper superparamagnetic 
dendrimers based on sites with unpaired spins. BL's are fundamentally one-dimensional (1D) 
systems, without any closed loops, that can be divided by cutting any bond. Their 
distinctive feature is exponential growth with g. Half of the sites for the BL with $Z = 3$ 
in Fig. \ref{fig1} are on the surface in generation g. BL's are bipartite: all nearest neighbor (NN) 
bonds in Fig.\ref{fig1} are between sites in even and odd $g$ that define two 
sublattices with different numbers of sites. The BL has open boundary conditions, 
a focal point at $g=0$ and an arbitrarily large boundary $g$. General theorems apply for 
the spectrum free electrons \cite{r8r} with NN transfer $t$ in a BL or for spins \cite{r8r1} 
with NN Heisenberg exchange $J > 0$. Three parameters generate a rich variety of $\rm BL(b,Z,g)$. 
In dynamical mean field theory \cite{r9}, the BL density of states with infinite 
connectivity is used as an initial guess for the density of states in higher dimensions. 
BL’s are models for strongly correlated systems \cite{r91}, alloys \cite{r92} 
and disordered systems \cite{r93}. The present study addresses antiferromagnetic 
(AF) Heisenberg exchange $J > 0$ in the BL with $b = Z = 3$

\begin{eqnarray}
H(g)&=& \sum_{\langle i,j \rangle } J \vec{s_i}\cdot \vec{s_j} 
\label{eq1}
\end{eqnarray}

\noindent The sum $\langle i,j \rangle $ is over all NN for either $s = 1/2$ or $s = 1$ sites. 
BL-Ising models limited to $s^z_is^z_j  $ interactions have been used 
to study finite spin glasses \cite{r10}. The electronic 
properties of correlated BL models still pose many challenges.\\

Recent advances in numerical techniques and computational resources 
have been applied to BL models. Methods include exact diagonalization (ED), 
quantum Monte Carlo (QCM) \cite{r11} and Density Matrix 
Renormalization Group (DMRG) \cite{r12}. DMRG is particularly well suited for 
1D systems such as Hubbard or extended Hubbard models, $t-J$ models and 
Heisenberg or related spin models. DMRG yields accurate properties for the 
ground state (gs) or low-energy excited states \cite{r13,r13r,r13r1,r13r2}. The DMRG challenge 
for a BL is the large number of surface sites in Fig. \ref{fig1}. 
Otsuka \cite{r14} applied DMRG to the BL, Eq. \ref{eq1}, with $s = 1/2$ sites and 
axially anisotropic NN exchange (XXZ model). Friedman \cite{r15} presented another DMRG 
algorithm for Eq. \ref{eq1} with $s = 1/2$. Lepetit et al. \cite{r16} used essentially 
the same algorithm to treat the Hubbard model version of Eq. \ref{eq1} that reduces to $J = 4t^2/U$ 
for $s = 1/2$ when the on-site repulsion $U$ is large compared NN electron transfer $t$. 
They also solved analytically the $ \rm H\ddot{u}ckel$ or tight-binding model with $U = 0$.\\ 

DMRG is a truncation procedure in which 
insignificant degrees of freedom of the system block, the right or left block in a chain, 
are discarded at each step with increasing system size. In 1D chains, the superblock 
consists of two blocks with dimension $m$ and two sites with $p$ degrees of freedom. The superblock dimension 
is $m^2 \times p^2$, with $p = 2s + 1$ in spin systems. The superblock of $\rm BL(3,3,g)$ has three 
blocks and hence goes as $m^3$. More branches $b$ increases the computational requirements. 
Otsuka \cite{r14} used four blocks and two new sites for $\rm BL(3,3,g)$ with $p = 2$ (s = 1/2). 
His superblock increased as $m^4 \times p^2$. The Friedman algorithm \cite{r15} with $b = 3$ and four new sites 
yields a superblock dimension of $m^3 \times p^4$. The DMRG algorithm in Section II has a 
superblock dimension $m^3 \times p$ or, more generally, $m^b \times p$, 
that makes $s=1$ sites accessible.\\

The paper is organized as follows. Section II presents and tests the new algorithm. 
Section III reports results for the BL in Fig. \ref{fig1} and Eq. \ref{eq1} 
up to $g = 11$ for $s = 1/2$ and 1 sites, and up to $g>20$ for $\langle s^z_0 \rangle $ at the 
focal point. We obtain the gs energy per site, 
the energy gap $\Delta$ that governs the gs magnetization and 
gs expectation values of $s^z_n$ for $n \le g$. We find spin correlation 
functions between the focal point and other sites and their convergence 
with increasing $g$. Section IV relates the DMRG results to a simple 
analytical model with localized states, to previous DMRG studies and 
to the question of long-range order in the infinite BL.
\begin{figure}[h]
\begin{center}
\includegraphics[height=9.0cm,width=9.0cm,angle=0]{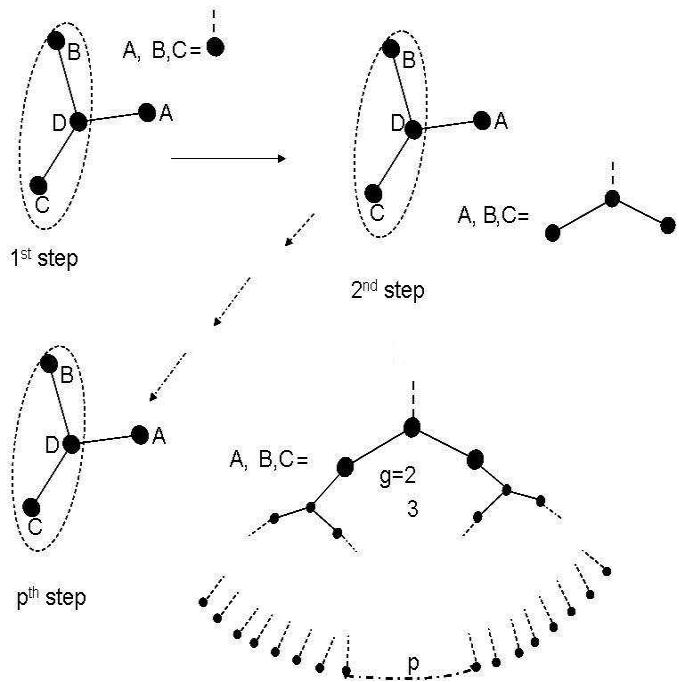}
\caption{ Schematic representation of BL growth from $g = 1$ at step 1 to $g = p$ at step $p$. 
The blocks A = B = C at step $g$ are given by Eq. \ref{eq2}. The focal point D is the new site added at each step.}
\label{fig2}
\end{center}
\end{figure}

\section{DMRG algorithm}
In this section, we present a new DMRG algorithm for the 
BL in Fig. \ref{fig1} with $b = Z = 3$. The principal change is 
how the lattice is grown. The total number of sites in $\rm BL(3,3,g)$ is

\begin{eqnarray}
N_T(g)=1+3(2^g-1)
\label{eq2}
\end{eqnarray}
\noindent The first step in Fig. \ref{fig2} contains four sites. Sites A, B and C are 
blocks each of whose size is $2^{g -1}$ at generation $g$. The focal point D is 
the new site added at each step. As shown in Fig. \ref{fig2}, each block 
A = B = C contains three sites in the second step for $g = 2$. 
Growth at step $g$ = 2, 3, ... is schematically represented as
\begin{eqnarray}
\setlength{\unitlength}{1mm}
\begin{picture}(60,16)
  \put(34,10){$D_g$}
  \put(8,7){$A_{g+1}=$}
  \put(28,6){\line(5,4){4}} 
  \put(24,4){$A_g$}
  \put(43,6){\line(-5,4){4}} 
  \put(44,4){$A_g$}
\end{picture}
\label{eq3}
\end{eqnarray}

\noindent $D_g$ is connected to the focal point D at step $g+1$.
The next step gives blocks of 7 sites at $g = 3$, and so on up to the BL with $N_T(g)$ sites. 
The procedure for BL(3,3,$g$) holds for other $b$ and $Z$.\\

The infinite DMRG algorithm for the BL proceeds along largely standard lines \cite{r12}: 
1. Start with the superblock matrix of four sites and find the eigenvalues of H. 2. Use the 
eigenvectors of the superblock to construct the density matrix of the 
new blocks $A_{g+1}$, initially for $g = 1$. Keep the eigenvectors of the $m$ 
largest eigenvalues, with $m$ chosen as discussed below. The density matrix dimension is $m^2 \times p$, 
where $m$ and $p$ refer to the block and degrees of freedom of the new site. 
Full diagonalization of the density matrix is carried out separately for large $m$ 
in sectors with different total $S^z$. 3. Renormalize the Hamiltonian of 
the new blocks and the operators that are necessary for the next step. 
These steps follow conventional DMRG \cite{r12}. 4. Construct the next ($g$ + 1) 
superblock from the three renormalized blocks $A_{g+1}$ and the new site. Diagonalize 
the matrix and retain the $m$ lowest eigenvalues and eigenvectors. Repeat steps 
2-4 until the desired system size is reached.\\
\begin{table}
\caption{The difference $\delta X$ between ED and DMRG with increasing $m$ for BL(3,3,3) 
with s=1/2 sites, where $X$ is the gs energy per site $\epsilon_0$, excitation energy $\Delta $,
the boundary spin $\langle s^z_3 \rangle $ and spin  correlation $\langle s^z_1s^z_3 \rangle $}
\begin{center}
\begin{tabular}{ccccc} \hline
$m$ ~~& ~~~$\delta \epsilon \times 10^9 $ &~~~ $ \delta \Delta \times 10^4 $ & ~~~ 
$\delta \langle s^z_3 \rangle \times 10^4    $ & ~~~$\delta  \langle s^z_0s^z_3 \rangle \times 10^5 $ \\ 
 &  &  & &  \\\hline 
10 &2670.8 & 42.593 & 89.48 & -1.193 \\
20 &560.08 & 41.441 & 0.030 & -0.722 \\
30 &552.99 & 41.212 & 0.031 & -0.801 \\
50 &0.001  & 0.0008 & 0.003 & -0.0001\\\hline
\end{tabular}
\label{tb1}
\end{center}
\end{table}

The superblock dimension of $m^3 \times p$ makes possible larger $m$ 
that increases the accuracy and larger $p = 2s + 1$. We can use $m = 60$ 
without much computational effort and find $10^{-13}$ or less for the weight of the 
discarded eigenvalues of blocks. DMRG is a variational method. 
Energies and correlation functions for given size $g$ converge better for 
finite DMRG \cite{r12}. We followed the standard approach of sweeping back and 
forth through different blocks. Care has to be taken in designing the finite DMRG algorithm 
due to the BL's complex structure. \\

We constructed the density matrix with equal weight for the lowest two eigenstates. 
As a first test of accuracy, we performed ED on BL(3,3,3) with 22 sites 
s=1/2 and on BL(3,3,2) with 10 sites s=1. There are $2^{22}$ and $3^{10}$ spin states, respectively.
DMRG results with increasing $m$ must eventually converge to ED. The evolution of the gs energy per site, 
$\delta \epsilon_0= \epsilon_0(m)- \epsilon_0(ED)$, with $m$ is shown in Table \ref{tb1} 
for the 22 sites systems. Also shown are the evolution of $ \delta \langle s^z_3 \rangle $ 
and the spin correlation $\delta \langle s^z_0s^z_3 \rangle $ between the focal point and 
the boundary. DMRG with $m=50$ is quantitative here. The Friedman algorithm whose super 
block increases as $m^3 \times p^4$ is limited to $m\approx 30$ and returns \cite{r15} 
$\delta \epsilon_0=2.2\times 10^{-6}$ with $m=29$ for infinite DMRG, which is comparable 
to $m=10$ in Table \ref{tb1}. DMRG with increasing $m$ also agrees quantitatively with ED for s=1 sites. \\
\begin{figure}[h]
\includegraphics[scale=0.25,height=9.0cm,width=7.0cm,angle=-90]{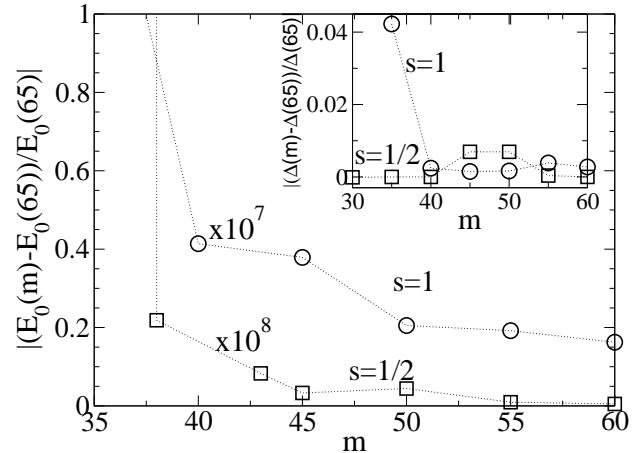}
\caption{ Ground state energy $E_0$ of spins 1/2 and 1 in a large BL(3,3,10) 
as a function of $m$ relative to $m$ = 65. The inset shows the $m$-dependence 
of the energy gap $\Delta$ to the ground state with $S^z = S(10) + 1$}
\label{fig3}
\end{figure}
 
We chose $m = 60$ on the basis of BL calculations for large systems with $g = 10$. Figure \ref{fig3} 
shows the evolution of $\epsilon_0(m)/\epsilon_0(65)$ with $m$ for s = 1/2 and 1 sites. Although s=1 
converges an order of magnitude more slowly, $m>40$ is adequate in either case. Preliminary results indicate
still slower convergence for s=3/2 sites. The inset of Fig. \ref{fig3} shows the slower, and not the monotonic, 
evolution of energy gap $\Delta(m)/\Delta(65)$. This is not unexpected since $\Delta $ is the difference between 
two extensive quantities.\\ 

We followed the first excited state in $S^z = S(g)$ with similar results. We kept $m = 60$ and 
did 8 sweeps of finite DMRG for the results in Section III. We estimate that the gs energy, spin 
densities and spin correlation functions are accurate to 4-5 
decimal places in larger systems, while energy gaps are accurate to 2-3 places.

\section{Results for BL with s = 1/2 and 1}
\begin{figure}[h]
\begin{center}
\includegraphics[scale=0.25,height=9.00cm,width=8.0cm,angle=-90]{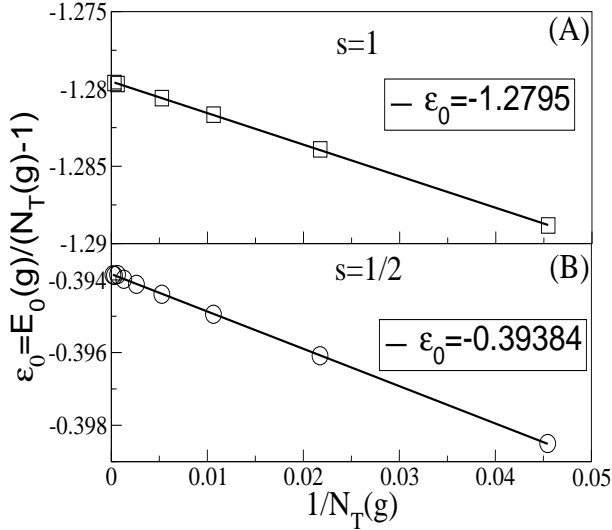}
\caption{ Ground state energy $\epsilon_0$ per bond of the BL, Eq. \ref{eq1}, 
with s = 1/2 or 1 sites, $g \leq 11$ generations and $N_T(g)$ sites. Linear 
extrapolation gives $\epsilon_0$ of the extended BL.}
\label{fig4}
\end{center}
\end{figure}
\begin{figure}[h]
\begin{center}
\includegraphics[scale=0.25,height=9.00cm,width=7.0cm,angle=-90]{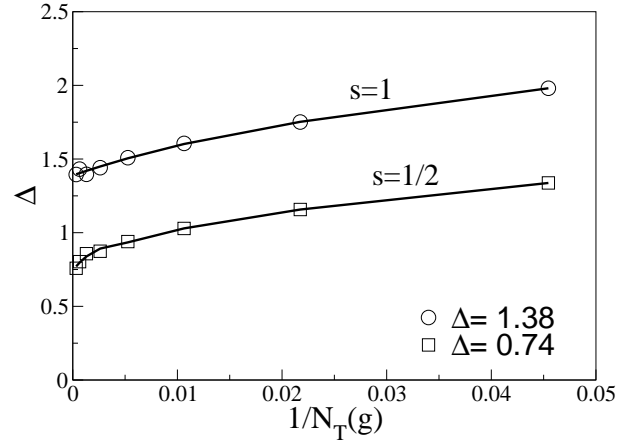}
\caption{Size dependence of the energy gap $\Delta$ to $S^z$ = S(g) + 1 for the BL, Eq. \ref{eq1}, 
with s = 1/2 or 1 sites, $g \leq 11$ and $N_T(g)$ sites.}
\label{fig5}
\end{center}
\end{figure}

In general, a BL of $N_T$ sites has $N_T - 1$ bonds, since only the focal 
point is not connected to a site with lower $g$. The gs energy per bond of BL(3,3,$g$) is
\begin{eqnarray}
\epsilon_0(g)=E_0(g)/(N_T(g)-1) 
\label{eq4}
\end{eqnarray}
\noindent where $E_0$ is the gs energy and $N_T$ is given in Eq. \ref{eq2}. 
Fig. \ref{fig5} shows $\epsilon_0(g)$ for $s = 1/2$ and 1 sites up to $g$ = 10. 
There is very little size dependence. The extended $s = 1/2$ and 1 systems have 
$\epsilon_0$=-0.39384 and -1.2795, respectively.\\

The gs has total $z$ component of spin $S^z(g) = 2^g s$. Eq. \ref{eq1} 
conserves $S$ and either a half-filled band or Heisenberg exchange yields $S(g) = 2^g s$. The gap $\Delta(g)$ 
to the gs in the $S^z = S(g) + 1$ sector governs the gs magnetization. 
The evolution of $\Delta(g)$ with $N_T(g)$ is shown in Fig. \ref{fig5}. The substantial gap of 
the extended system is discussed in Section IV. We also found doubly degenerate excitations $\Delta'(g)$, 
comparable to $\Delta(g)$, in the $S^z(g)$ sector for both $s = 1/2$ and 1. The $C_3$ symmetry of 
the BL leads naturally to $E$ states. Since we compute $S^z$ rather than $S$, the gs automatically appears also 
in sectors with $S^z < S^z(g)$. The second and third excited states for $S^z$ = $S^z(g) - 1$ decrease with 
$N_T(g)$ and vanish in the extended system within our numerical accuracy.\\

\begin{figure}[h]
\includegraphics[scale=0.25,height=9.00cm,width=7.0cm,angle=-90]{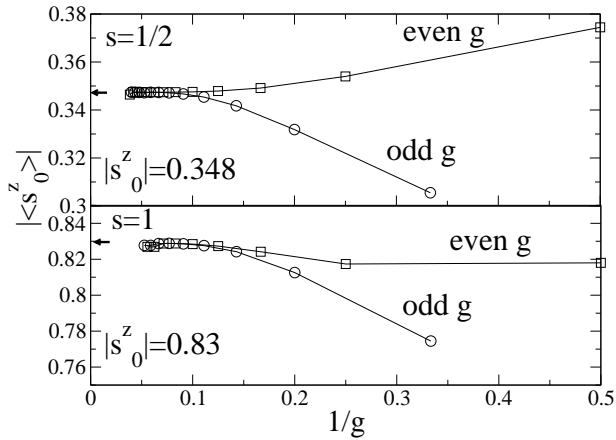}
\caption{  Magnitude of $\langle s^z_0 \rangle$ at the focal point of a BL up 
to $g$ = 26 generations for s = 1/2 and up to $g$ = 24 for s = 1 sites. 
Even and odd $g$ form separate series that merge at large $g$.}  
\label{fig6}
\end{figure}

The gs expectation values of $ \langle s^z_p \rangle $ in generation p are listed in Table \ref{tb2} 
for the BL with $g$ = 10. The $g$ = 9 and 11 results are almost the same. The gs 
has long-range order (LRO) that corresponds to staggered magnetization in successive generations. 
The largest $\langle s^z_p \rangle $ is at the boundary, $p = g$. The smallest magnitude is at $p = g - 1$  
next to the boundary, and $|\langle s^z_n \rangle |$ near the focal point become equal for large $g$. 
The convergence of $ |\langle s^z_0 \rangle |$ to 0.348 is shown in Fig. \ref{fig6} up to $g = 26$, 
a huge BL of $3 \times 2^{26}$ sites, and agrees with the previous estimate of 0.35 \cite{r16}. 
The $s = 1$ limit up the $g = 24$ is $| \langle s^z_0 \rangle |$ = 0.83, and the staggered magnetization of 
both gs becomes constant near the focal point for $g > 10$. By contrast, a half-filled BL of free electron has 
$\langle s^z_0 \rangle=2/(g+1)$ for odd $g$ and $\langle s^z_0 \rangle =0$ for even $g$ 
or in the extended system. \\ 
Table \ref{tb2} also lists the gs 
expectation values of $ \langle \vec{s}_{p-1} \cdot \vec{s}_p \rangle $ in successive generations of BL(3,3,10) 
with $s = 1/2$ and 1 sites. The $g$ = 9 and 11 values are similar. Variations of 
$ \langle \vec{s}_{p-1} \cdot \vec{s}_p \rangle $ at $g$ and $g-1$ are reduced near the focal point. 
The BL-Ising model has $ \langle s^z_p \rangle  = (-1)^ps$ and $\langle s^z_{p-1} s^z_p\rangle = -s^2$. 
Table \ref{tb2} shows that $s = 1$ is closer to the Ising model than $s = 1/2$, where quantum fluctuations are larger.\\

\begin{figure}[h]
\includegraphics[scale=0.25,height=9.00cm,width=7.0cm,angle=-90]{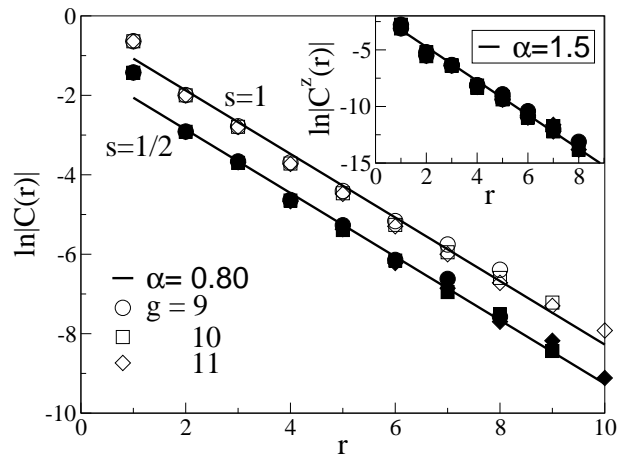}
\caption{ Spin correlation functions $C(r)$ in Eq. \ref{eq6} between the focal 
point and generation $r \leq g$ in BL's with $g$ = 9, 10, 11 and s = 1/2 or 1 sites. 
The inset shows the z component, $C^z(r)$. }
\label{fig7}
\end{figure}

We define radial spin correlation functions as 
\begin{eqnarray}
C(r)& = & (-1)^r (\langle \vec{s_0} \cdot \vec{s_r} \rangle - \langle 
s^z_0 \rangle \langle s^z_r  \rangle)
\label{eq6}
\end{eqnarray}
\noindent with $r = 1,2...g$. The mean-field contribution is explicitly excluded. As seen in 
Fig. \ref{fig7}, $C(r)$ decreases as $\rm exp(-\alpha r)$ with $\alpha = 0.80$ for both 
$s = 1/2$ and 1. Accurate DMRG makes it possible to compute small $C(r)$ up to $r \approx 10$. 
The inset shows the related correlation function $C^z(r)$ with $ \langle s^z_0 s^z_r  \rangle $ instead of the dot
\begin{table}[h]
\begin{center}
\caption { Ground state expectation values of $\langle s^z_p \rangle$ and $\langle \vec{s}_{p-1} \cdot \vec{s_{p}} \rangle$ 
for s=1/2 and 1 site in BL(3,3,10). The focal point and boundary are $p=0$ and $p=10$ respectively.}
\begin{tabular}{ccccc} \hline
p & \multicolumn{2}{c}{s=1/2} & \multicolumn{2}{c}{s=1} \\

   & $~~~\langle s^z_p \rangle~~~ $  & $~~~\langle \vec{s}_{p-1} \cdot \vec{s_{p}}  \rangle ~~~$ 
$~~~~$ & $~~~\langle s^z_p \rangle ~~~$ & $~~~\langle \vec{s}_{p-1} \cdot \vec{s_{p}} \rangle ~~~$  \\\hline

0  &  0.347 & --     & 0.829 &  --  \\ 
1  & -0.345 &-0.359 &-0.828 & -1.214 \\
2  &  0.347 &-0.361 & 0.828 & -1.214\\
3  & -0.342 &-0.357 &-0.826 & -1.213\\
4  &  0.348 &-0.364 & 0.826 & -1.215\\
5  & -0.332 &-0.350 &-0.817 & -1.208\\
6  &  0.348 &-0.371 & 0.820 & -1.218\\
7  & -0.307 &-0.334 &-0.789 & -1.191\\
8  &  0.356 &-0.392 & 0.816 & -1.238\\
9  & -0.249 &-0.289 &-0.691 & -1.104\\
10 &  0.393 &-0.456 & 0.872 & -1.397 \\\hline
\end{tabular}
\label{tb2}
\end{center}
\end{table}
 product, 
which decreases even faster. $C(r)$ is almost completely due to transverse spin components. Since the number of boundary sites goes as
$n(g)=1.5 \rm exp(gln2)$, we have $n(g)C(g) \rightarrow 0 $ for large $g$. \\ 

The gs degeneracy is lifted when an applied magnetic field $h$ is added to Eq. \ref{eq1}. 
The lowest Zeeman level goes as $-hS(g)$. With increasing $h$, the gs becomes the lowest Zeeman level 
of a state with $S^z > S(g)$, an excited state at $h = 0$. When $S^z = S(g) + 1$, the crossover 
\begin{figure}[h]
\includegraphics[scale=0.25,height=9.00cm,width=7.5cm,angle=-90]{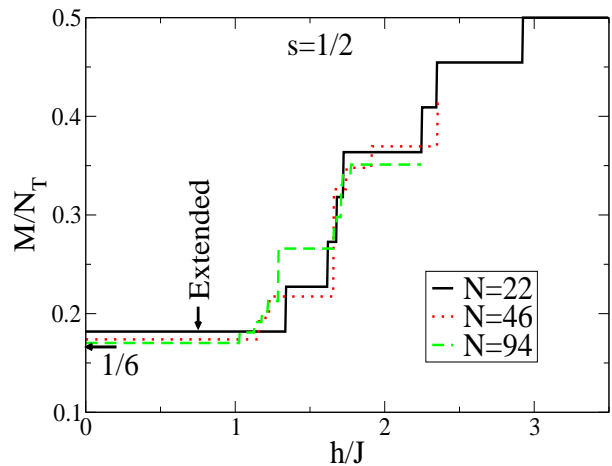}
\caption{ Exact magnetization $M$ per site of BL(3,3,3) with s = 1/2 sites 
as a function of applied field $h$. The $g$ = 4 and 5 magnetization is 
DMRG up to finite $h$. The extended BL has an $M = 1/6$ plateau up to $h=\Delta$, 
the gap shown in Fig. \ref{fig5}.}
\label{fig8}
\end{figure}
field is related to the zero-field energy in the two sectors,
\begin{eqnarray}
h=\Delta(g)=[E_0(S(g)+1)-E_0]/J.
\label{eq7}
\end{eqnarray}
 The first crossover may be to a state with higher $S^z$, when Eq. \ref{eq7} 
has a multiple of $h$. The gs magnetization per $s = 1/2$ site of BL(3,3,$g$) 
is shown in Fig. \ref{fig8} for $g =$ 3, 4 and 5. The first jump is to $S^z$ = $S(g)$ + 1. Complete alignment 
at large $h$ leads to $M = 1/2$ in reduced units. The extended BL has a magnetic gs with $M = 1/6$ at $h$ = 0 
and an initial increase at $\Delta$ marked with an arrow in Fig. \ref{fig8}.  
\section{Discussion}
DMRG results for $H(g)$ in Eq. \ref{eq1} are similar for BL’s with $s = 1/2$ and 1 sites in Fig. \ref{fig1}. 
The gs has $S(g) = 2^gs$ as expected on general grounds. The larger sublattice has $N_A(2) = 2^{g+1} - 1$ 
sites and the smaller has $ N_B(g) = 2^g - 1$ sites. The difference $N_A(g) - N_B(g)$ is $2^g$. As seen in Table \ref{tb2}, 
the gs has staggered magnetization with LRO and AF spin correlations in successive generations. 
Spin correlation functions $C(r)$ in Fig. \ref{fig8} decay rapidly and exponentially. 
Magnetization has a substantial gap $\Delta(g)$ that remains finite in the extended s = 1/2 or 1 system.\\ 

We interpret these results in terms of a simple analytical approximation. We partition $H(g)$ with $J = 1$ in Eq. \ref{eq1} 
as $H = H_0 + V$,  

\begin{eqnarray}
H_0&=&N_B(g)h(3) \nonumber\\
h(3)&= &\vec{s_2}\cdot (\vec{s_1}+\vec{s_3}) 
\label{eq8}
\end{eqnarray}

\noindent The trimer $h(3)$ is elementary and $H_0$ accounts for exactly 2/3 of the exchanges. For example, the outermost two generations of BL(3,3,5) in Fig. \ref{fig1} contain eight trimers per arm. 
Each $s = 1/2$ trimer has a gs with $s = 1/2$ at $e_0= -1$, another doublet at $e = 0$ 
and a quartet, $s = 3/2$, at $e = 1/2$. The gs of an $s = 1$ trimer is $e_0 = -3$. 
The perturbation $V$ contains an extra site $q$ in the larger sublattice and remaining exchanges. 
Each of $N_A(q)$ choices of $q$ uniquely defines 
the sites of $N_B(g)$ trimers. The middle and end 
sites are necessarily in the smaller and larger sublattice, respectively, and all exchanges in $V$ 
are between a middle and end site. There is only one kind of trimer-trimer or trimer-$q$ interaction.\\

The gs energy of $H_0$ for $s = 1/2$ and 1 sites is $-N_B(g)$ and $-3N_B(g)$, respectively. Spin correlation 
functions of adjacent sites for $s = 1/2$ and 1 are

\begin{eqnarray}
\langle \vec{s_1}\cdot \vec{s_2} \rangle=-1/2, -3/2.
\label{eq9}
\end{eqnarray}

\noindent The spin densities of the central and terminal sites for s = 1/2 are

\begin{eqnarray}
2\langle s^z_2 \rangle=-1/3 &,& 2\langle s^z_1 \rangle=2/3,
\label{eq10}
\end{eqnarray}

\noindent The corresponding spin densities for $s = 1$ are $-1$ in the middle and 3/2 at the ends. 
The gs of $H_0$ is a product over trimers and the spin at $q$,

\begin{eqnarray}
|\Psi^{(0)}(q,g)\rangle &=& |\alpha_q \rangle \prod^{N_B(g)}_{j=1}|\psi_j \rangle 
\label{eq11}
\end{eqnarray}

\noindent First-order perturbation theory in V lifts the degeneracy between $q$ in generations $g$ and $< g$.\\

Site $q$ is a $1/N_T$ correction for large g. The energy per site of the infinite BL is, to first order in V, 
\begin{eqnarray}
\langle \Psi^{(0)}|H|\Psi^{(0)}\rangle/3N_B(g) &=& \epsilon_0(s)/3+\langle s^z_1 \rangle \langle s^z_2 \rangle/3
\label{eq11}
\end{eqnarray}

\noindent Parallel spins give lower energy due to the opposite signs of spin densities. 
The $s = 1/2$ result is -19/54 = –0.35185 per site in first order and -0.39082 in 
second order in $V$, very close the DMRG result of 0.39385 in Fig. \ref{fig4}. 
The $s = 1$ result is -9/8 = –1.125 per site in first order and -1.3152 in second order, 
slightly below -1.2796 in DMRG. The variational theorem holds for the energy in first order, 
but not in second order. The first order energy is lowest for parallel spins of $N_B(g)$ trimers 
and site $q$ that properly gives $S^z = S(q)s$. Moreover, $\Psi^{(0)}(q,g)$ immediately rationalizes 
low-energy spin flips in states with $S < S(g)$ for either $s = 1/2$ or 1 sites.\\

Site $q$ is a NN of one middle site when $q$ is in generation $g$ and of three 
middle sites otherwise. 
The second term of Eq. \ref{eq12} is more negative for $q \leq g - 2$ than for $q = g$ 
by 2$\langle s^z_2 \rangle (s- \langle s^z_1 \rangle) $ for $s$ = 1/2 or 1. To first order in $V$, 
site $q$ in not on the boundary and  $g$, $g - 1$ in Fig. \ref{fig1} 
are trimers with end sites $g$. $H_0$ accounts for all exchanges between 
generations $g$ and $g - 1$ while $V$ contains all exchanges between $g - 1$ and $g - 2$. 
The trimer approximation leads to $\langle s^z_g \rangle = 1/3$ and $ 
\langle s^z_{g-1} \rangle = -1/6$, 
compared to 0.40 and -0.25 for DMRG for $s = 1/2$ sites in Table \ref{tb2}. 
The correlation functions are $\langle \vec{s}_{g-1} \cdot \vec{s}_g \rangle = -1/2$ compared to -0.44 for DMRG. 
Trimers have reduced $\langle \vec{s}_{g-2} \cdot \vec{s}_{g-1}\rangle = -1/18$, well below the DMRG result of 
-0.30 but consistent with reduced correlation in $g - 1$ and $g - 2$. The $s = 1$ 
BL has $\langle s^z_g \rangle = 3/4$ and $\langle s^z_{g-1} \rangle = -1/2$ 
in the trimer approximation and 0.87, -0.69 in DMRG (Table 2). Trimers have 
$\langle \vec{s}_{g-1} \cdot \vec{s}_g \rangle =-3/2$ for s=1 while DMRG gives -1.397.\\

A trimer of $s=1/2$ sites must be excited to a quartet state $s = 3/2$ with excitation energy 3/2 under $H_0$ to obtain $S^z = S(g) + 1$. 
One of the $N_B(g)$ trimers in $\Psi^{(0)}(q,g)$ is changed to $|\phi\rangle = |\alpha \alpha \alpha \rangle$. 
A normalized function  with a quartet is

\begin{eqnarray}
|\Phi(g) \rangle &=&(N_B(g))^{-1/2} \sum_{m=1}^{N_B(g)} |\phi_m \rangle |\Psi^{(0)}(q,g)\rangle/|\psi_m \rangle\nonumber\\
                & & 
\label{eq12}
\end{eqnarray}

\noindent The quartet is delocalized over the BL by the $s^+s^-$ terms of $V$. To first order in $V$, 
the excitation energy to $S = S(g) + 1$ is
\begin{eqnarray}
\langle \Phi |H|\Phi\rangle -\langle \Psi^{(0)}|H| \Psi^{(0)} \rangle  & = & \Delta^{(1)}=1.0
\label{eq13}
\end{eqnarray}

 \noindent Delocalization lowers the energy by 2/3 while the diagonal $s^zs^z$ contribution raises 
the energy by 1/6. The net effect is to lower the excitation from 3/2 to 1 for $s = 1/2$, 
somewhat above  $\Delta= 0.74$ for the extended BL in Fig. \ref{fig5}.\\

The sharp distinction between NN exchanges in $H_0$ and $V$ for the outermost 
three generations is lost in the interior. $H_0$ contains trimers that spans 
three generations when $q \leq g - 2$. When $q \ne 0$, the central site in Fig. \ref{fig1} 
is the middle site of a trimer for even $g$ and the end site for odd $g$. Although trimers 
imply  intermediate spin density and spin correlations near the focal point, 
there are variations between even and odd $g$ in contrast to identical $ \langle s^z_0 \rangle $ 
in Fig. \ref{fig6}. Similarly, $C(r)$ in Eq. \ref{eq6} for $\Psi^{(0)}(q,g)$ is strictly 
limited to $r$ = 1 or 2 since trimers span at most span three generations. The  
function $\Psi^{(0)}(q,g)$ is localized, more localized than the DMRG gs, but it 
rationalizes DMRG results in some detail.\\

DMRG results for NN exchange $J$ in Eq. \ref{eq1} for s = 1/2 and 1 sites 
in BL(3,3,g) are closely similar, in sharp contrast to the fundamentally different 
behavior of 1D chains of s = 1/2 and 1 sites \cite{r17}. The s = 1 chain with additional 
NN terms $J(s_i \cdot s_j)^2/3$ in Eq. \ref{eq1} is a valence bond solid (VBS) with 
rigorously known gs properties \cite{r18}. The VBS on BL(3,3,g) has s = 3/2 site, yet another 
added term to Eq. \ref{eq1}, and $\langle s^z_0 \rangle = 0$ and hence no LRO in the extended system \cite{r18}.
Both the spin and Hamiltonian of the VBS are different, and no DMRG has been performed on that system. 
For Eq. 1, DMRG indicates a gs with staggered magnetization and finite $\langle s^z_0 \rangle $  
in Fig. \ref{fig6} at the focal point for either s = 1/2 or 1. The gs has LRO and short-range spin 
correlations $C(r)$ in Eq. \ref{eq6} that decrease exponentially and more rapidly 
than the number of boundary sites.

\section{Conclusions}
The DMRG algorithm in Section II makes it possible to treat 
the BL in Fig. \ref{fig1} with s = 1 sites. It improves the accuracy 
for s = 1/2 sites and yields low-energy excitations. All DMRG results for either s=1/2 or 1 sites be understood qualitatively
in terms of a trimer model for BL(3,3,g). \\ 

Preliminary results for BL(3,3,g) with s = 3/2 sites are satisfactory. The addition of a focal point with $p$ 
degrees of freedom at each step can be used for BL's with more arms $b > 3$ or higher 
connectivity $Z > 3$, although with steep increases of computational resources. 
The bottleneck is the dimension $m^{Z-1} \times p$ of the block whose density matrix 
is constructed at each step; $m \approx 70$ for $Z = 3$ becomes $m \approx 17$ for $Z = 4$, 
which has limited accuracy. A comparable reduction to $m \approx 20-30$ limited previous algorithms \cite{r14,r15,r16} for 
BL(3,3,g) to s=1/2 sites. The dimension $m^b \times p$ of the superblock is less 
serious because only a few eigenvalues are required at each step. \\

{\bf Acknowledgments.}  MK thanks SR White for discussion and BJ Topham for reading the manuscript carefully.
We thank the National Science Foundation for partial support of this work
through the Princeton MRSEC (DMR-0819860).
 
\end{document}